# Near-infrared excitation of nitrogen-doped ultrananocrystalline diamond photoelectrodes in saline solution

Andre Chambers,[1] Arman Ahnood,[1,*] Samira Falahatdoost,[1] Steve Yianni,[2] David Hoxley,[2] Brett C. Johnson,[3] David J. Garrett,[1] Snjezana Tomljenovic-Hanic,[1] and Steven Prawer[1]

[1]School of Physics, University of Melbourne, Melbourne, Victoria 3010, Australia

[2]Department of Chemistry and Physics, La Trobe University, Bundoora, Victoria 3086, Australia

[3]Centre for Quantum Computation & Communication Technology, School of Physics, University of Melbourne, VIC 3010, Australia

*Nitrogen-doped ultrananocrystalline diamond (N-UNCD) is a promising material for a variety of neural interfacing applications, due to its unique combination of high conductivity, bioinertness, and durability. One emerging application for N-UNCD is as a photoelectrode material for high-precision optical neural stimulation. This may be used for the treatment of neurological disorders and for implantable bionic devices such as cochlear ear implants and retinal prostheses. N-UNCD is a well-suited material for stimulation photoelectrodes, exhibiting a photocurrent response to light at visible wavelengths with a high charge injection density [A. Ahnood, A. N. Simonov, J. S. Laird, M. I. Maturana, K. Ganesan, A. Stacey, M. R. Ibbotson, L. Spiccia, and S. Prawer, Appl. Phys. Lett. **108**, 104103 (2016)]. In this study, the photoresponse of N-UNCD to near-infrared (NIR) irradiation is measured. NIR light has greater optical penetration through tissue than visible wavelengths, opening the possibility to stimulate previously inaccessible target cells. It is found that N-UNCD exhibits a photoresponsivity which diminishes rapidly with increasing wavelength and is attributed to transitions between mid-gap states and the conduction band tail associated with the graphitic phase present at the grain boundaries. Oxygen surface termination on the diamond films provides further enhancement of the injected charge per photon, compared to as-grown or hydrogen terminated surfaces. Based on the measured injected charge density, we estimate that the generated photocurrent of oxygen terminated N-UNCD is sufficient to achieve extracellular stimulation of brain tissue within the safe optical exposure limit.*

The use of light-based techniques for neural stimulation is an area of growing interest, with potential applications in the treatment of neurological disorders and in implantable bionic devices [1–4]. In particular, optically-driven electrodes have the potential to offer wireless stimulation with much greater spatial resolution than conventional electrically-driven electrodes [2]. This approach relies on the transduction of light into electrical signals in order to stimulate neural tissue, caused by the separation of photo-excited charge carriers in a semiconducting electrode [2]. Materials such as photoconductive silicon [5–8], conductive polymers [9–11], and quantum dots [12–14] have been extensively studied for this purpose. However, these photoactive surfaces have often been found to exhibit limited biostability or produce cytotoxic reactions [15–20].

Diamond is a material with the potential to address these issues, due to its well-known durability and biocompatibility [21–25]. Single crystal diamond is a wide-gap semiconductor with an intrinsic photoresponse band at unsafe ultraviolet frequencies and hence is not useful for neural stimulation applications. In contrast, nitrogen-doped ultrananocrystalline diamond (N-UNCD) is highly conductive due to the presence of $sp^2$ bonded carbon at the diamond grain boundaries, and has been shown to exhibit a photoresponse at much longer wavelengths [26,27], making it a material well-suited for photostimulation [28]. In addition, the surface chemistry of N-UNCD may be altered to exhibit high electrochemical capacitance, a desirable attribute for neuromodulation electrodes [23,29].

The potential of N-UNCD for use as a photoelectrode material has been previously investigated, with the finding that it exhibits a photoresponse to wavelengths of 450 nm or shorter, meeting the requirements for extracellular and intercellular stimulation within the safe optical exposure limit [28]. In the present study, the feasibility of extending the wavelength range of the photoresponse is investigated. Longer wavelengths have greater optical penetration depth in biological tissue, and reduce the potential for phototoxic effects resulting in higher safe optical exposure limits [19,30]. To test this, the spectral response of N-UNCD is measured and analysed with reference to the known band structure. The effect of the surface chemistry on the electrochemical capacitance and charge transfer mechanisms was also examined. Finally, the capability of this technique to achieve threshold charge injection for the stimulation of neurons is evaluated, taking into account various parameters such as cell type, laser pulse parameters, and the size of the stimulating electrodes.

The N-UNCD thin-film samples used in this study were grown in an *Iplas* microwave plasma-assisted CVD system on polycrystalline diamond (PCD) and nanodiamond-seeded silicon substrates. Films were grown to a thickness of approximately 30 μm. Details of the N-UNCD seeding and deposition processes have been reported elsewhere [23]. Samples underwent further processing to terminate the surface in various ways. Oxygen plasma was applied for 16 hours at a power of 50 W and a pressure of 0.6 mbar to achieve oxygen surface termination. For hydrogen surface termination, samples were treated in a hydrogen plasma for 3 minutes at a substrate temperature of 800°C, 1200 W power and a pressure of 100 mbar. The surface topology and conductivity of the N-UNCD films were assessed by conductive atomic force microscopy (C-AFM), using an *Asylum Research MFP-3D* system under contact mode (*Asylum Research 'Econo-SCM-PIC'*, resonance frequency 13 kHz, force constant 0.2 N/m).

The photoresponse of the N-UNCD electrodes was investigated through photoelectrochemical measurements undertaken in a three-electrode cell connected to a potentiostat (*Gamry*). For high temporal resolution photocurrent measurements, an oscilloscope (*LeCroy*) was used in conjunction with a pre-amplifier (*DLPCA-200, FEMTO*), while for spectral measurements a lock-in amplifier (*Stanford Research Systems*) was utilised to record the average photocurrent amplitude. The electrochemical cell consisted of a custom-made chamber containing 0.15 M NaCl solution, with a Pt disk counter electrode and Ag/AgCl reference electrode (*eDAQ*). The working electrode was the N-UNCD sample, placed underneath an opening in the bottom of the electrochemical chamber and sealed with a 3 mm diameter O-ring (see Fig. 1(a)). A bias voltage of approximately 15 mV was applied between the working and counter electrodes to achieve zero current (open circuit condition) prior to illumination.

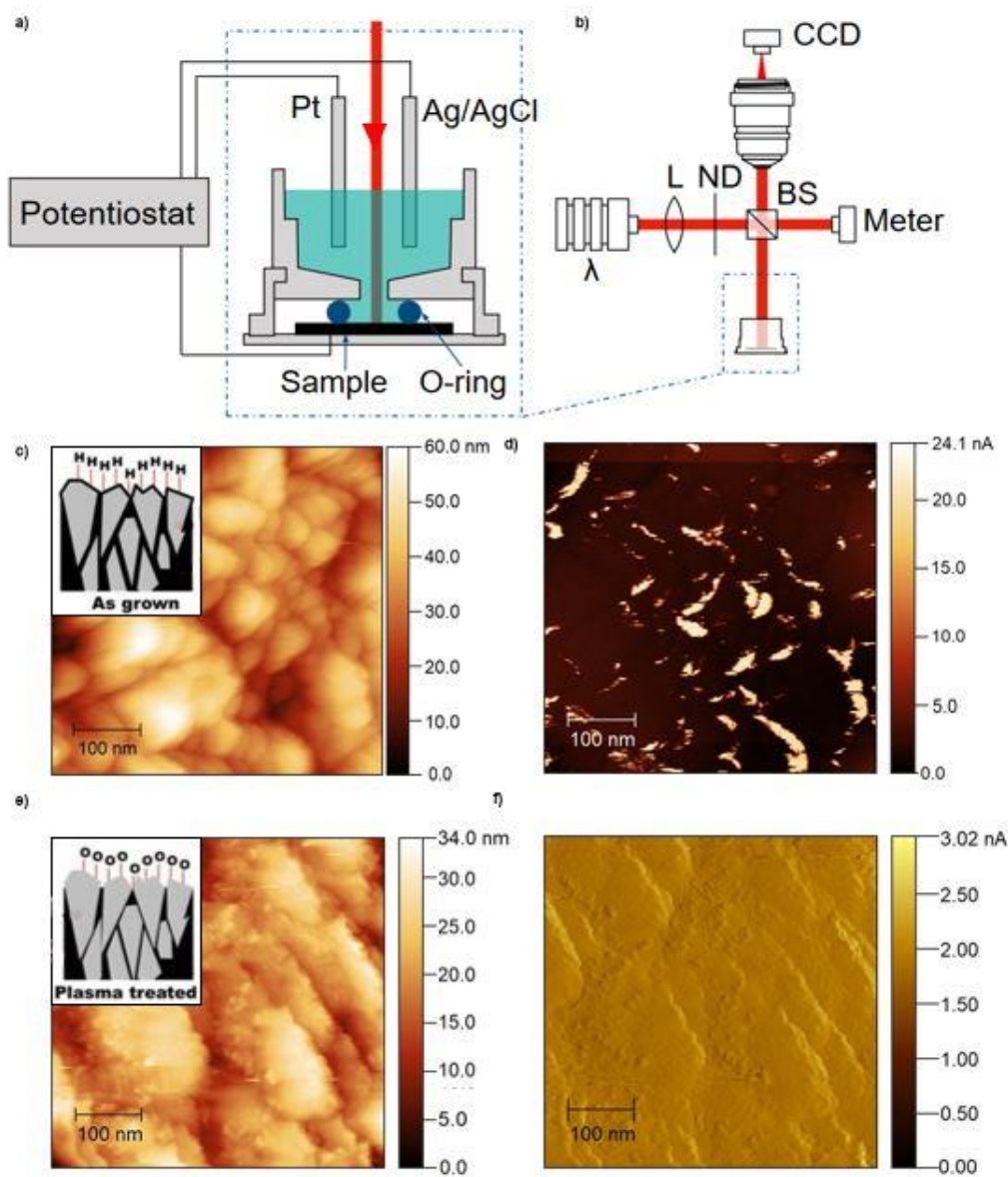

**Fig. 1: (a)** Illustration of the electrochemistry chamber, indicating the placement of the working (sample), counter (Pt) and reference (Ag/AgCl) electrodes. **(b)** Schematic of the optical setup, showing the optical paths to the sample, imaging system, and optical power meter. **(c)** Topographic and **(d)** conductive maps of the same surface area of as-grown N-UNCD obtained by C-AFM. **(e)** Topographic and **(f)** conductive maps of the surface of oxygen terminated N-UNCD shows that the plasma treatment removes the conductive hot-spots. Shown inset is a diagram of the proposed etching of the graphitic phase at the surface after oxygen plasma treatment.

The light source used for photoelectrochemical measurements was a 1.6 Watt 808 nm near infrared (NIR) laser diode (*Wuhan Lilly Electronics*), which was directed through a compound lens to a 50/50 non-polarising beamsplitter, and then onto the sample (Fig. 1(b)). The second beam transmitted through the beamsplitter was directed to an optical power meter (*Coherent FieldMaxII-TOP*). The lenses were mounted on an adjustable stage to allow alteration of the optical path length and thus the laser spot size incident on the sample. The laser spot size was directly measured using a calibrated microscope camera mounted above the sample. The sample was also illuminated using this setup for a range of other wavelengths by means of mounted LEDs (*Thorlabs, RS Electronics*).

N-UNCD is a mixed phase diamond-based material with high grain boundary content. As displayed in C-AFM topographic map (Fig. 1(c)), the as-grown N-UNCD film surface is composed of clusters of nanodiamond grains separated by grain boundary regions. The conductivity map of the same surface region (Fig. 1(d)) shows that these grain boundary regions are highly conductive, consistent with a high graphitic $sp^2$ phase content [31]. The C-AFM measurement on an oxygen terminated N-UNCD film (Fig. 1(e), 1(f)) indicates that plasma treatment eliminates the conductive channels, suggesting preferential etching of the graphitic phase at the grain boundary [28].

The chemical termination of the diamond surface leads to a dramatic different in the film's electrochemical properties. This is evident in cyclic voltammetry measurements, whereby a triangular voltage ramp is externally applied to the sample and the current response is measured. As shown in Fig. 2(a), as-grown and hydrogen terminated N-UNCD samples exhibit similar electrochemical behaviour, with both voltammograms having limited hysteresis suggesting low electrochemical capacitance. The voltage limits at which solvent breakdown begins to occur (water window) are also similar for as-grown and hydrogen terminated samples. The small peak around -0.1 V for the as-grown sample is attributed to the reactivity of a surface rich in carbon species [32]. In contrast, the voltammogram for the oxygen terminated sample exhibits a much greater hysteresis compared to the previous two samples, indicating a larger electrochemical capacitance. This capacitance was found to be 1170 ± 240 $\mu F\ cm^{-2}$ in agreement with previous work [23]. This is a very high value for diamond, and is comparable to other commonly used electrode materials such as titanium nitride [23,33]. The reason for the increased capacitance of oxygen terminated N-UNCD has previously been suggested to be a combination of factors, including the increased surface area, etching of the conductive grain boundary regions, and the positive electron affinity of the oxygen terminated surface [23].

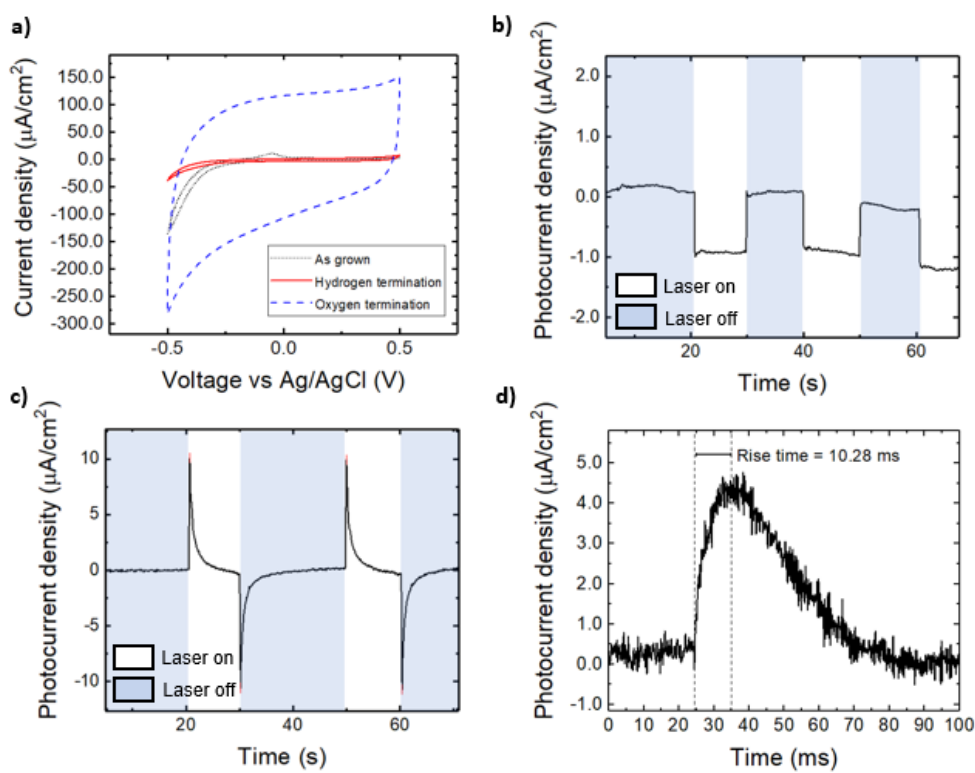

**Fig. 2: (a) Cyclic voltammograms of different N-UNCD surface terminations performed at 100 mV/s in dark conditions with an exposed electrode area of 0.031 cm$^2$. Results indicate that oxygen termination gives rise to a significant increase in the photoelectrode capacitance. (b) Transient photocurrent of hydrogen terminated N-UNCD and (c) oxygen terminated N-UNCD in response to pulsed illumination at 808 nm with a maximum intensity of 9.55 W mm$^{-2}$ and illuminated area of 0.26 mm$^2$. In contrast to hydrogen termination, capacitive coupling between the oxygen terminated photoelectrode and saline solution is evident. (d) High temporal resolution photoelectrochemical measurement on oxygen terminated N-UNCD showing the rise time of 10.3 ms.**

The different electrochemical properties of hydrogen and oxygen terminated N-UNCD also affect the photocurrent response in solution. As shown in Fig. 2(b), the photocurrent plot of the hydrogen terminated sample displays a continuous current flow in response to illumination. The as-grown sample exhibits a similar response. This behaviour is consistent with a Faradaic charge transfer mechanism, where electrons are directly injected into the electrolyte [29,34]. This could be due to the negative electron affinity of the hydrogen terminated surface, making it energetically favourable for photogenerated electrons to pass through the N-UNCD-electrolyte interface into solution, in line with previous works [28,35,36]. The Faradaic mechanism of charge transfer is not ideal for neural stimulation applications, due to the formation of reactive oxygen species (ROS) which may lead to tissue damage and electrode degradation [28,29]. On the other hand, the surface of oxygen terminated N-UNCD possesses a positive electron affinity, with the photocurrent plot displaying a capacitive charging and discharging in response to light (Fig 2(c)) [28]. Unlike Faradaic charge transfer, this capacitive mechanism does not directly inject charge into the electrolyte but instead indirectly causes a redistribution of ions within solution, and thus produces a redistribution of charge [34]. This mechanism does not lead to the production of ROS, and is more favourable for neural stimulation [34]. It was determined that the photocurrent peaks for oxygen terminated N-UNCD have an average rise time of $10.3 \pm 1.8$ ms (Fig. 2(d)). The large difference in the photocurrent fall times shown in Fig. 2(c) and 2(d) could be explained by the built-in impedance of the potentiostat circuit employed in the first measurement compared to the oscilloscope circuit used in the second. Moreover, it was observed that the photoresponse of oxygen terminated N-UNCD is opposite in polarity to hydrogen terminated N-UNCD. Oxygen termination also produces a larger photocurrent by approximately one order of magnitude, which has previously been attributed to the formation of dense oxygen terminated diamond nanocrystals at the surface [28]. This magnitude of the photocurrent is also comparable to other photoactive materials previously studied for neural stimulation [1,10].

Another important feature of the photoresponse is the wavelength of the incident light, in this case in the NIR range (808 nm). To further investigate the dependence on wavelength, the transient photoresponse of oxygen terminated N-UNCD was measured over a range of wavelengths from 375 nm to 810 nm. The photoresponse was defined as a normalised quantity corresponding to the peak photocurrent per incident photon rate. As shown in Fig. 3(a), the photoresponse of N-UNCD sharply decreases above 400 nm, in agreement with previous work [27]. There is also a small peak in the photoresponse around 730 nm. This behaviour has been previously observed as persistent photocurrents after UV illumination of CVD-grown diamond thin films [37], and has been attributed to a deep-level defect associated with nitrogen clusters [38]. For N-UNCD, we propose that the same defect is accessible without the need for UV activation due to the enhanced tunnelling between nanometre sized diamond grains and graphitic matrix. While this peak suggests that incident light at a wavelength of 730 nm produces the optimum photoresponse in the NIR range, 808 nm light still possesses a greater optical penetration depth and lower potential for photothermal damage [39]. Therefore, 808 nm light was investigated further to determine whether threshold charge injection for neural stimulation could be produced in N-UNCD within the safe NIR exposure limits.

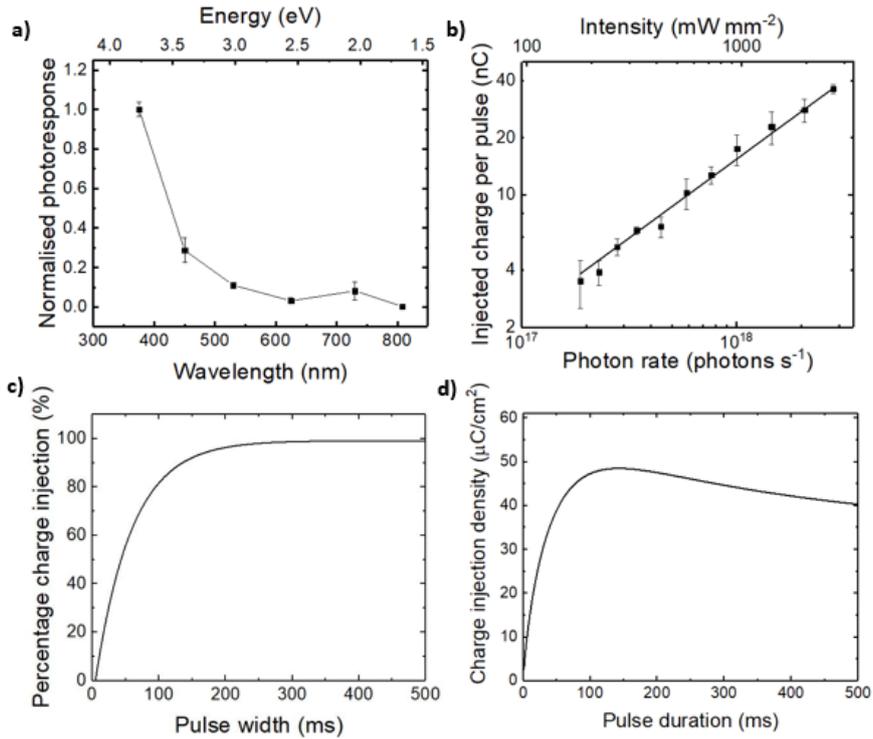

Fig. 3: (a) The normalised photoresponse (photocurrent per unit photon rate) of oxygen terminated N-UNCD as a function of wavelength. The photoresponse at 810 nm is 0.23% of that at 375 nm. Average photocurrent values are obtained using a lock-in amplifier as described in the text. (b) Injected charge versus photon rate for oxygen terminated N-UNCD at 808 nm wavelength, determined by integrating the area under each photocurrent pulse. (c) Percentage of maximum charge injection of oxygen terminated N-UNCD as a function of the optical pulse duration. (d) Charge injection of N-UNCD modelled as a function of the optical pulse duration of 808 nm light, at the safe optical exposure limit using the data in (b) and (c) as input. Based on this relation, it was determined that a 25 μm diameter N-UNCD electrode would generate a sufficient photoresponse for *in vivo* stimulation in response to 808 nm within the safe optical exposure limit.

There are several mechanisms through which a sub-bandgap photoresponse could occur. In the NIR range of wavelengths, the most likely mechanisms include mid-gap absorption due to defects and two-photon absorption [40]. This latter process occurs when two photons are simultaneously absorbed, leading to an energy transition corresponding to the sum of their energies [41]. Typically, two-photon absorption only becomes the dominant absorption process at high optical intensities, and can be identified by a quadratic relationship between the optical absorption and incident photon rate [41].

The relationship between the photoresponse and light intensity was determined at an incident wavelength of 808 nm, as shown in Fig. 3(b). The gradient of this logarithmic plot (0.84 ± 0.02) indicates that the curve is approximately linear and not the quadratic relationship expected from two-photon absorption, suggesting that this is not the dominant mechanism of sub-bandgap photoexcitation. Instead, the observed sub-bandgap photoresponse is likely due to recombination at mid-gap defect states [42]. In particular, the photoresponse has been suggested to be due to mid-bandgap π and π* states associated with the graphitic grain boundary regions of N-UNCD [28,43]. As noted by Ahnood *et al.*, the π→π* transition of approximately 2.1-2.5 eV is consistent with their observed photoresponse at 450 nm [28]. However, the disordered nature of the graphitic content results in broadening of the defect states allowing photo-excitation at much longer wavelengths [43]. Moreover, this disorder introduces conduction band tail states which may contribute to the sub-bandgap photoresponsivity [26]. This is supported by previous works showing optical absorption in the NIR range for CVD nanodiamond films [44,45].

After the photoelectrochemical characterisation of the N-UNCD film, the feasibility of using this technique for neural stimulation was assessed. One factor which is critical in this assessment is the light intensity, which must be within the safe optical exposure limit for *in vivo* stimulation. High optical intensities of NIR light bear the risk of photothermal damage to the biological tissue [46]. For the range of pulse frequencies useful for neural stimulation, assuming a laser spot size less than 25 μm in diameter, the safe optical exposure limit for the retina is $6.93 \times 10^{-4}$ $C_T$ $t^{-0.25}$ Watts, where $C_T$ is a constant equal to $10^{0.002(\lambda-700)}$, t is the pulse duration in seconds, and λ is the wavelength in nanometres [47]. This evaluates to $1.14 \times 10^{-3}$ $t^{-0.25}$ Watts for incident 808 nm light, which corresponds to a local temperature rise at least ten times lower than the threshold for photothermal damage [47]. In the case of light incident on brain tissue rather than the retina, there are currently no widely accepted safe exposure standards [48]. However, it has been suggested that substantially higher intensities may be safely utilised at low pulse frequencies [4]. Nevertheless, for this analysis the above optical exposure limit has been chosen as a conservative estimate.

Within this safe-optical exposure limit, the charge injection density per pulse may be optimised through the choice of laser parameters, such as the optical intensity and pulse duration. Reducing the optical pulse duration may cut off the photocurrent pulse, and thus decrease the quantity of injected charge as shown in Fig. 3(c). To find the optimum laser parameters for *in vivo* neural stimulation, the injected charge per pulse at the safe optical exposure limit was modelled based on the relationship depicted Fig. 3(b) for a range of pulse durations, weighted by the function extracted from the data shown in Fig. 3(d). It was found that the maximum charge injection density of 48 μC cm$^{-2}$ was attainable with an optical pulse duration of 130 ms within the safe optical exposure limit. It was also determined that this amount of charge injection was not significantly affected by the temperature or oxygen concentration in solution (for details, see ref. [49]).

To determine whether this amount of injected charge is sufficient for cellular stimulation, it is necessary to consider a variety of parameters, including the type of cell, and the size and proximity of the stimulation electrode. It has been previously found that the threshold charge injection density for extracellular stimulation of retinal cells is 0.023 mC cm$^{-2}$ [50], while deep brain stimulation requires just 0.0023-0.0067 mC cm$^{-2}$ [51]. The threshold for intracellular stimulation is even lower, requiring a charge injection density of 0.001 mC cm$^{-2}$ [52]. For the stimulation method presented in this study, the charge injection density may be found by dividing the injected charge per pulse by the surface area of the N-UNCD electrode. Assuming that the photoresponse depends only on the incident photon rate and is independent of the electrode size, a circular N-UNCD electrode with the same dimensions as the laser spot (25 μm in diameter) would generate sufficient charge injection density for the extracellular stimulation of brain neurons in response to optical pulse durations of 5 ms. This assessment does not take into account the effect of local thermal gradients, which have been shown to contribute to membrane depolarisation of neurons in response to infrared light [4].

In summary, the performance of N-UNCD as an optically-driven electrode for neural stimulation applications was investigated over a range of wavelengths. It was determined that N-UNCD exhibits a sub-bandgap photoresponse which diminishes with increasing wavelength, an effect attributed to recombination at mid-gap defect states associated with the graphitic grain boundary regions. It was also found that oxygen surface termination enhances the photoresponse by promoting a capacitive charge transfer mechanism. The photosensitivity of N-UNCD in the NIR range will enable greater optical penetration in biological tissue than shorter wavelengths. This provides a further benefit in addition to its high bioinertness and durability, in comparison to photoactive materials such as photoconductive silicon, conductive polymers and quantum dots [20]. Finally, it was predicted that *in vivo* stimulation of brain tissue is feasible using oxygen terminated N-UNCD electrodes driven by 5 ms 808 nm light pulses within the safe optical exposure limit. These findings further advance N-UNCD as a well-suited material candidate for optically-driven stimulation electrodes.

ACKNOWLEDGEMENTS

This work was performed in part at the Australian National Fabrication Facility (ANFF), a company established under the National Collaborative Research Infrastructure Strategy, through the La Trobe University Centre for Materials and Surface Science. The authors also wish to acknowledge the technical assistance of Dr. Matias Maturana at Clinical Sciences, Department of Medicine, University of Melbourne. This research was supported by the Australian Research Council, through Linkage Grant LP160101052. D.J.G. is supported by a National Health and Medical Research Council of Australia Project Grant GNT1101717. A.A. gratefully acknowledges the William Stone Trust funding, University of Melbourne.

STATEMENT OF INTERESTS

S.P. is a shareholder, director and chief technology officer of iBIONICS, a company developing a diamond based retinal implant. D.J.G. and S.P. are directors and shareholders in Carbon Cybernetics, a company developing a carbon based brain machine interface. The remaining authors declare that the research was conducted in the absence of any commercial or financial relationships that could be construed as a potential conflict of interest.

FOOTNOTES AND REFERENCE CITATIONS

* arman.ahnood@unimelb.edu.au

[1]   L. Bareket, N. Waiskopf, D. Rand, G. Lubin, M. David-Pur, J. Ben-Dov, S. Roy, C. Eleftheriou, E. Sernagor, O. Cheshnovsky, U. Banin, and Y. Hanein, Nano Lett. **14**, 6685 (2014).

[2]   F. Di Maria, F. Lodola, E. Zucchetti, F. Benfenati, and G. Lanzani, Chem. Soc. Rev.

**47**, 4757 (2018).

[3] L. Fenno, O. Yizhar, and K. Deisseroth, Annu. Rev. Neurosci. **34**, 389 (2011).

[4] C. Paviolo, A. C. Thompson, J. Yong, W. G. A. Brown, and P. R. Stoddart, J. Neural Eng. **11**, 065002 (2014).

[5] P. Fromherz and A. Stett, Phys. Rev. Lett. **75**, 1670 (1995).

[6] Y. Goda and M. A. Colicos, Nat. Protoc. **1**, 461 (2006).

[7] A. Starovoytov, J. Choi, and H. S. Seung, J. Neurophysiol. **93**, 1090 (2005).

[8] J. Suzurikawa, M. Nakao, Y. Jimbo, R. Kanzaki, and H. Takahashi, IEEE Trans. Biomed. Eng. **56**, 2660 (2009).

[9] V. Gautam, M. Bag, and K. S. Narayan, J. Am. Chem. Soc. **133**, 17942 (2011).

[10] D. Ghezzi, M. R. Antognazza, M. Dal Maschio, E. Lanzarini, F. Benfenati, and G. Lanzani, Nat. Commun. **2**, 166 (2011).

[11] M. R. Antognazza, M. Di Paolo, D. Ghezzi, M. Mete, S. Di Marco, J. F. Maya-Vetencourt, R. Maccarone, A. Desii, F. Di Fonzo, M. Bramini, A. Russo, L. Laudato, I. Donelli, M. Cilli, G. Freddi, G. Pertile, G. Lanzani, S. Bisti, and F. Benfenati, Adv. Healthc. Mater. **5**, 2271 (2016).

[12] K. Lugo, X. Miao, F. Rieke, and L. Y. Lin, Biomed. Opt. Express **3**, 447 (2012).

[13] T. C. Pappas, W. M. S. Wickramanyake, E. Jan, M. Motamedi, M. Brodwick, and N. A. Kotov, Nano Lett. **7**, 513 (2007).

[14] E. Molokanova, J. A. Bartel, W. Zhao, I. Naasani, M. J. Ignatius, J. A. Treadway, and A. Savtchenko, Invit. Corp. 1 (2008).

[15] H. Yamato, M. Ohwa, and W. Wernet, Electrochim. Acta **397**, 163 (1995).

[16] S. Marciniak, X. Crispin, K. Uvdal, M. Trzcinski, J. Birgerson, L. Groenendaal, F. Louwet, and W. R. Salaneck, Synth. Met. **141**, 67 (2004).

[17] R. Biran, D. C. Martin, and P. A. Tresco, Exp. Neurol. **195**, 115 (2005).

[18] R. Hardman, Environ. Health Perspect. **114**, 165 (2006).

[19] A. L. Vahrmeijer, Nat Rev Clin Oncol **18**, 1199 (2013).

[20] L. Bareket-Keren and Y. Hanein, Int. J. Nanomedicine **9**, 65 (2014).

[21] M. Amaral, A. G. Dias, P. S. Gomes, M. A. Lopes, R. F. Silva, J. D. Santos, and M. H. Fernandes, J. Biomed. Mater. Res. - Part A **87**, 91 (2008).

[22] P. Bajaj, D. Akin, A. Gupta, D. Sherman, B. Shi, O. Auciello, and R. Bashir, Biomed. Microdevices **9**, 787 (2007).

[23] W. Tong, K. Fox, A. Zamani, A. M. Turnley, K. Ganesan, A. Ahnood, R. Cicione, H. Meffin, S. Prawer, A. Stacey, and D. J. Garrett, Biomaterials **104**, 32 (2016).

[24] C. Popov, W. Kulisch, J. P. Reithmaier, T. Dostalova, M. Jelinek, N. Anspach, and C. Hammann, Diam. Relat. Mater. **16**, 735 (2007).

[25] T. Lechleitner, F. Klauser, T. Seppi, J. Lechner, P. Jennings, P. Perco, B. Mayer, D.

Steinmüller-Nethl, J. Preiner, P. Hinterdorfer, M. Hermann, E. Bertel, K. Pfaller, and W. Pfaller, Biomaterials **29**, 4275 (2008).

[26] P. Achatz, J. A. Garrido, M. Stutzmann, O. A. Williams, D. M. Gruen, A. Kromka, and D. Steinmüller, Appl. Phys. Lett. **88**, 101908 (2006).

[27] K. J. Pérez Quintero, S. Antipov, A. V. Sumant, C. Jing, and S. V. Baryshev, Appl. Phys. Lett. **105**, 123103 (2014).

[28] A. Ahnood, A. N. Simonov, J. S. Laird, M. I. Maturana, K. Ganesan, A. Stacey, M. R. Ibbotson, L. Spiccia, and S. Prawer, Appl. Phys. Lett. **108**, 104103 (2016).

[29] D. J. Garrett, K. Ganesan, A. Stacey, K. Fox, H. Meffin, and S. Prawer, J. Neural Eng. **9**, 016002 (2012).

[30] W.-F. Cheong, S. A. Prahl, and A. J. Welch, IEEE J. Quantum Electron. **26**, 2166 (1990).

[31] M. Mertens, I.-N. Lin, D. Manoharan, A. Moeinian, K. Brühne, and H. J. Fecht, AIP Adv. **7**, 015312 (2017).

[32] Y. Tzeng, S. Yeh, W. C. Fang, and Y. Chu, Sci. Rep. **4**, 4531 (2015).

[33] J. D. Weiland, D. J. Anderson, and M. S. Humayun, IEEE Trans. Biomed. Eng. **49**, 1574 (2002).

[34] D. R. Merrill, M. Bikson, and J. G. R. Jefferys, J. Neurosci. Methods **141**, 171 (2005).

[35] L. Zhang and R. J. Hamers, Diam. Relat. Mater. **78**, 24 (2017).

[36] D. Zhu, L. Zhang, R. E. Ruther, and R. J. Hamers, Nat Mater **12**, 836 (2013).

[37] E. Rohrer, C. Graeff, C. Nebel, M. Stutzmann, H. Guttler, and R. Zachai, Mater. Sci. Eng. **B46**, 115 (1997).

[38] C. E. Nebel, A. Waltenspiel, M. Stutzmann, M. Paul, and L. Schäfer, Diam. Relat. Mater. **9**, 404 (2000).

[39] E. A. Boettner and J. R. Wolter, Invest. Ophthalmol. Vis. Sci. **1**, 776 (1962).

[40] M. Casalino, G. Coppola, M. Iodice, I. Rendina, and L. Sirleto, Sensors **10**, 10571 (2010).

[41] V. Bredikhin, M. Galanin, and V. N. Genkin, Sov. Phys. USP **110**, 3 (1973).

[42] Y. V. Pelskov, A. Y. Sakharova, M. D. Krotova, L. L. Bouilov, and B. V. Spitsyn, J. Electroanal. Chem. **228**, 19 (1987).

[43] Y. Wang, M. Jasiswal, M. Lin, S. Saha, B. Ozyilmaz, and K. P. Loh, Elsevier **6**, 1018 (2012).

[44] M. Nesládek, K. Meykens, L. Stals, M. Vaněček, and J. Rosa, Phys. Rev. B - Condens. Matter Mater. Phys. **54**, 5552 (1996).

[45] O. Williams, M. Nesladek, S. Michaelson, A. Hoffman, E. Osawa, K. Haenen, and R. B. Jackman, Diam. Relat. Mater. **17**, 1080 (2008).

[46] A. N. Chester, S. Martellucci, A. M. Verga Scheggi, and North Atlantic Treaty Organization. Scientific Affairs Division., *Laser Systems for Photobiology and*

*Photomedicine* (Springer US, 1991).

[47] F. C. Delori, R. H. Webb, and D. H. Sliney, J. Opt. Soc. Am. A **24**, 1250 (2007).

[48] G. Strangman, D. A. Boas, and J. P. Sutton, Biol. Psychiatry **52**, 679 (2002).

[49] See Supplemental Material at [URL will be inserted by publisher] for the dependence of the photocurrent on temperature and oxygen concentration.

[50] Y. Yamauchi, L. M. Franco, D. J. Jackson, J. F. Naber, R. O. Ziv, J. F. Rizzo, H. J. Kaplan, and V. Enzmann, J. Neural Eng. **2**, S48 (2005).

[51] A. M. Kuncel and W. M. Grill, Clin. Neurophysiol. **115**, 2431 (2004).

[52] J. T. Robinson, M. Jorgolli, A. K. Shalek, M. H. Yoon, R. S. Gertner, and H. Park, Nat. Nanotechnol. **7**, 180 (2012).